\documentclass[sigconf,edbt]{acmart-edbt2024}

\def\BibTeX{{\rm B\kern-.05em{\sc i\kern-.025em b}\kern-.08em
    T\kern-.1667em\lower.7ex\hbox{E}\kern-.125emX}}

\usepackage{booktabs} 
\usepackage{ragged2e}

\usepackage[a-3b,mathxmp]{pdfx}

\usepackage{graphicx}
\usepackage{balance}  
\usepackage{multirow}
\usepackage{url}
\usepackage{caption}
\usepackage{listings}
\usepackage{enumitem}
\usepackage{siunitx}
\usepackage{adjustbox}
\usepackage{comment}
\usepackage{wrapfig}
\usepackage[frozencache,cachedir=.]{minted}
\usepackage[breakable]{tcolorbox}
\definecolor{stateflow-pink}{HTML}{EFC9DB}
\definecolor{stateflow-orange}{HTML}{FFDD8C}
\definecolor{stateflow-blue}{HTML}{14A3D4}
\definecolor{stateflow-gray}{HTML}{F3F3F3}

\usepackage[ruled]{algorithm2e}

\newcommand{\para}[1]{\vspace{1mm}\noindent\textbf{#1.}}

\newcommand\blfootnote[1]{%
  \begingroup
  \renewcommand\thefootnote{}\footnote{#1}%
  \addtocounter{footnote}{-1}%
  \endgroup
}

\setcopyright{rightsretained}

\acmDOI{}

\acmISBN{978-3-89318-091-2}

\acmConference[EDBT 2024]{27th International Conference on Extending Database Technology (EDBT)}{25th March-28th March, 2024}{Paestum, Italy}
\acmYear{2024}

\AtBeginDocument{%
  \providecommand\BibTeX{{%
    \normalfont B\kern-0.5em{\scshape i\kern-0.25em b}\kern-0.8em\TeX}}}

\definecolor{n1_code}{HTML}{405063}
\definecolor{green}{HTML}{008800}
\definecolor{props}{HTML}{db597c}
\definecolor{nicerblue}{HTML}{0066bb}
\definecolor{deepblue}{HTML}{0000dd}
\definecolor{velvetred}{HTML}{bb0066}

\definecolor{gray}{rgb}{0.5,0.5,0.5}
\definecolor{lg}{HTML}{DDDDDD}
\definecolor{CC}{HTML}{666666}
\definecolor{numbers}{HTML}{444444}

\definecolor{bgreen}{HTML}{20BC5F}

\lstdefinelanguage{java}{
  morekeywords={abstract,case,catch,class,def,%
    do,else,extends,false,final,finally,toMatrix,%
    for,if,implicit,import,match,mixin,%
    new,null,object,override,package,%
    private,protected,requires,return,sealed,%
    super,this,throw,trait,true,try,%
    type,val,var,while,with,yield
    },  
  otherkeywords={=>,<-,<\%,<:,>:,\#,@},
  sensitive=true,
  morecomment=[l]{//},
  morecomment=[n]{/*}{*/},
  morestring=[b]",
  morestring=[b]',
  morestring=[b]"""
}

\lstdefinelanguage{python}{
  morekeywords={abstract,case,catch,class,def,%
    do,else,extends,false,final,finally,toMatrix,%
    for,if,implicit,import,match,mixin,%
    new,null,object,override,package,%
    private,protected,requires,return,sealed,%
    super,this,throw,trait,true,try,%
    type,val,var,while,with,yield,stateflow,self,bool,int
    }, 
  otherkeywords={=>,<-,<\%,<:,>:,\#,@},
  sensitive=true,
  morecomment=[l]{//},
  morecomment=[n]{/*}{*/},
  morestring=[b]",
  morestring=[b]',
  morestring=[b]"""
}

\lstdefinestyle{javalang}{
  frame=none,
  language=java,
  showstringspaces=false,
  columns=flexible,
  basicstyle={\footnotesize\ttfamily},
  numbers=none,
  numberstyle=\tiny\color{numbers},
  numbersep=5pt,
  keywordstyle=\bfseries\color{n1_code},
  commentstyle=\color{CC},
  stringstyle=\bfseries,
  breaklines=false,
  breakatwhitespace=false,
  tabsize=2,
  xleftmargin=.0in,
  captionpos=b,
  keepspaces=true,
  escapechar=|
}

\lstdefinestyle{pythonlang}{
  frame=single,
  language=Python,
  showstringspaces=false,
  emph={User,__init__, set_balance, add_to_basket, __key__, buy_item, buy_item_0, buy_item_1},
  emphstyle={\bfseries\color{velvetred}},
  columns=flexible,
  basicstyle={\footnotesize\ttfamily},
  identifierstyle=\color{nicerblue},
  numbers=left,
  numberstyle=\tiny\color{deepblue},
  numbersep=5pt,
  keywordstyle=\bfseries\color{green},
  commentstyle=\color{CC},
  stringstyle=,
  breaklines=false,
  breakatwhitespace=false,
  tabsize=2,
  xleftmargin=.0in,
  captionpos=b,
  keepspaces=true,
  escapechar=|
}

\lstdefinestyle{javainline}{
  frame=none,
  language=java,
  showstringspaces=false,
  columns=flexible,
  basicstyle={\ttfamily},
  numbers=none,
  numberstyle=\tiny\color{gray},
  keywordstyle=\bfseries\color{n1_code},
  commentstyle=\color{gray},
  stringstyle=\bfseries,
  breaklines=false,
  breakatwhitespace=false,
  tabsize=2,
  xleftmargin=.1in,
  captionpos=b,
  keepspaces=true,
  escapechar=|,
  moredelim=**[is][\color{n1_code}]{`}{`},
  moredelim=**[is][\color{props}]{~}{~},
}

\newcommand{\sysname}{StateFlow}

\begin{document}

\title{Stateful Entities: Object-oriented Cloud Applications as \\ Distributed  Dataflows}

\author{Kyriakos Psarakis$^{*}$}
\affiliation{%
  \institution{Delft University of Technology}
  \city{Delft} 
  \state{Netherlands} 
}
\email{k.psarakis@tudelft.nl}

\author{Wouter Zorgdrager$^{*}$}
\affiliation{%
  \institution{Delivery Hero SE}
  \city{Berlin} 
  \state{Germany} 
}
\email{wouter.zorgdrager@deliveryhero.com}

\author{Marios Fragkoulis}
\affiliation{%
  \institution{Delivery Hero SE}
  \city{Berlin} 
  \state{Germany} 
}
\email{marios.fragkoulis@deliveryhero.com}

\author{Guido Salvaneschi}
\affiliation{%
  \institution{University of St. Gallen}
  \city{St. Gallen} 
  \state{Switzerland} 
}
\email{guido.salvaneschi@unisg.ch}

\author{Asterios Katsifodimos}
\affiliation{%
  \institution{Delft University of Technology}
  \city{Delft} 
  \state{Netherlands} 
}
\email{a.katsifodimos@tudelft.nl} 

\renewcommand{\shortauthors}{}

\begin{abstract}
Although the cloud has reached a state of robustness, the burden of using its resources falls on the shoulders of programmers who struggle to keep up with ever-growing cloud infrastructure services and abstractions. As a result, state management, scaling, operation, and failure management of scalable cloud applications, require disproportionately more effort than developing the applications' actual business logic.

Our vision aims to raise the abstraction level for programming scalable cloud applications by compiling \emph{stateful entities} --  a programming model enabling imperative transactional programs authored in Python --  into stateful streaming dataflows. We propose a compiler pipeline that analyzes the abstract syntax tree of stateful entities and transforms them into an intermediate representation based on stateful dataflow graphs. It then compiles that intermediate representation into different dataflow engines, leveraging their exactly-once message processing guarantees to prevent state or failure management primitives from ``leaking'' into the level of the programming model. Preliminary experiments with a proof of concept implementation show that despite program transformation and translation to dataflows, stateful entities can perform at sub-100ms latency even for transactional workloads.\blfootnote{$^*$Both authors contributed equally to this work.}
\end{abstract}

%
%



\maketitle

\section{Introduction}

Organizations nowadays enjoy reduced costs and higher reliability, but cloud developers still struggle to manage infrastructure abstractions that leak through, in the application layer. As a result, managing application components, such as service invocation, messaging, and state management, require much more effort than the development of the application's business logic \cite{blanas-transactions}. Worse, moving a  cloud application between cloud providers is prohibitive, due to significant differences in the underlying systems.

While there are multiple approaches for distributed application programming (e.g.,  Bloom \cite{alvaro2010boom}, Hilda \cite{yang2006hilda}, Cloudburst \cite{cloudburst}, AWS Lambda, Azure Durable Functions, and Orleans \cite{bykov2011orleans,bernstein2014orleans}), in practice developers mainly use libraries of popular general purpose languages such as Spring Boot in Java, and Flask in Python. 

None of these approaches offers message processing guarantees, failing to support \emph{exactly-once processing}: the ability of a system to reflect the changes of a message to the state exactly one time. Instead, they offer at-most- or at-least-once processing semantics. Programmers then have to  ``pollute'' their business logic with consistency checks, state rollbacks, timeouts, retries, and idempotency \cite{microservices-survey, microservices-drawbacks}.

We argue that no matter how we approach cloud programming unless an execution engine can offer exactly-once processing guarantees so that it can be assumed at the level of the programming model, we will never remove the burden of distributed systems aspects from programmers. To the best of our knowledge, the only systems able to guarantee exactly-once message processing \cite{silvestre2021clonos,CarboneEF17} at the time of writing, are batch \cite{dean2008mapreduce, dryadlinq, CUSTOM:web/Spark} and streaming \cite{storm, CarboneKE15, murray2013naiad} dataflow systems. However, their programming model follows the paradigm of functional dataflow APIs which are cumbersome to use and require training, and heavy rewrites of the typical imperative code that developers prefer to use for expressing application logic.

For these reasons, we argue that the dataflow model should be used as a low-level intermediate representation (IR) for the modeling and execution of distributed applications, but not as a programmer-facing model. Technically, one of the main challenges in adopting a dataflow-based IR is that the dataflow model is essentially functional, with immutable values being propagated across operators that typically do not share a global state. Hence, adopting a dataflow-based IR entails translating (arbitrary) imperative code into the functional style.
Compiler research has systematically explored only the opposite direction: to compile code in functional programming languages into a representation that is executable on imperative architectures -- like virtually all modern microprocessors. Yet, the translation from imperative to functional or dataflow programming remains largely unexplored.

\begin{figure*}[t]
    \centering
    \includegraphics[width=1\textwidth]{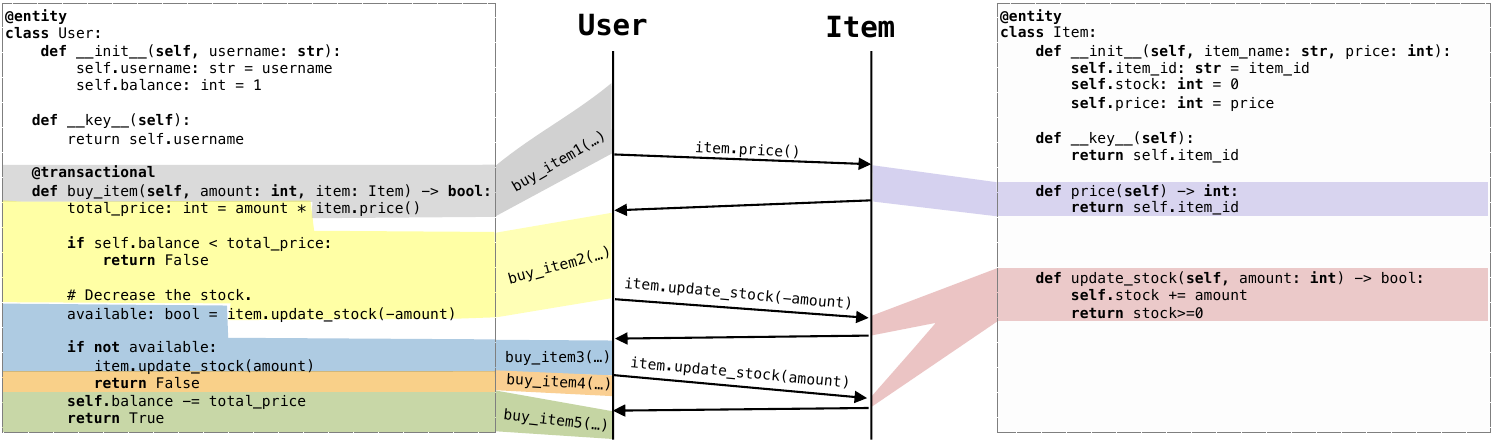}
    \caption{Two stateful entities: \texttt{User} and \texttt{Item}. The content of imperative functions is split into multiple functions that access the common state of a given entity. Those functions are then encoded into a stateful dataflow that can be executed in a distributed streaming dataflow engine. As a result, $i)$ imperative code is executed in an event-based manner without the need to block, and $ii)$ the code retains exactly-once processing guarantees without the need for programmers to write failure-handling code such as state management, call retries or idempotency.}
    \label{fig:calls}
\end{figure*}

This paper presents a prototypical programming model, compiler pipeline, and IR that compiles imperative, transactional object-oriented applications into distributed dataflow graphs and executes them on existing dataflow systems. 
Instead of designing an external Domain-Specific Language (DSL) for our needs, we opted for an internal DSL embedded in Python - a language that is already popular for cloud programming.
Specifically, a given Python program is first compiled into an IR, an enriched stateful dataflow graph that is independent of the target execution engine. That dataflow graph can then be compiled and deployed to a variety of distributed systems. The current set of supported systems includes Apache Flink Statefun and \sysname{} -- our own dataflow system built for the needs of such low-latency cloud applications. The choice of a runtime system is completely independent of the application layer, which allows switching to different runtime systems with no changes to the application code.

\vspace{2mm}
\noindent The contributions of this paper go as follows:
\begin{itemize}
    \item To the best of our knowledge, this is the first work to propose compiling and executing imperative programs into distributed, stateful streaming dataflows.
    \item We present a compiler pipeline that analyzes an object-oriented application and transforms it into an IR tailored to stateful dataflow systems.
    \item We describe an IR for cloud applications and how that IR translates to a dataflow execution graph, targeting a variety of distributed systems, thereby making cloud applications portable across different systems and infrastructures.
    \item We compare Stateflow, a novel transactional dataflow system, against Apache Flink Statefun and demonstrate the limitations of existing dataflow systems, motivating further research. Our experimental evaluation shows that Stateflow incurs low latency in the YCSB+T \cite{dey2014ycsb+} workload.
    \item Despite the promising early results, several research questions remain open. We detail those in this paper and lay out a future research agenda.
\end{itemize}

The proposed system presented in this paper can be found at:  \url{https://github.com/delftdata/stateflow}. A preliminary version of this paper is included as an abstract in CIDR 2023 \cite{entities-abstract}.

\section{From Imperative Code to Dataflows}

Historically, imperative programming and functional programming have evolved in parallel:  imperative as a direct codification of (operational) computational models (e.g., Von Neumann architecture, Turing machines), and functional inspired by mathematical abstractions (e.g., lambda calculus, program denotation).
While functional programming has been embraced by a number of languages (e.g., Haskell \cite{hudak1992report}, ML \cite{harper1986standard}) imperative programming has taken the scene, with most mainstream languages featuring object-oriented (mutable) abstractions.
Over the last years, imperative languages like Java and Python, which support a large variety of domain-specific packages, e.g., networking, statistics, numeric computation, etc. have become extremely popular among non-expert programmers.

Yet, the benefits of functional programming have been known for a while. Most notably, functional code is often {\it embarrassingly parallelizable} because of the lack of side effects and mutability. Developers working with imperative languages -- let alone non-expert developers -- can hardly access this feature.

\subsection{Approach Overview}

The main principle behind our compiler pipeline is that developers simply annotate Python classes with \texttt{@stateflow} and the system automatically analyzes and transforms these classes into an intermediate representation which is then transformed into stateful dataflow graphs, ready to be deployed on a dataflow system.  Similar to (Virtual) Actors \cite{bykov2011orleans,wyatt2013akka}, \emph{entities} can make calls to methods of other entities. \autoref{fig:calls} depicts two sample entities: \texttt{User} and \texttt{Item}. Details of the programming model are provided in \autoref{sec:assumptions}.

In the first pass of an Abstract Syntax Tree (AST) static analysis, we extract the class's variables (i.e. instance attributes referenced with \texttt{self}), the names of each method, and all respective types indicated by the programmer (\autoref{sec:assumptions}). In the second round of analysis, classes that interact with each other are identified in order to create a function call graph (\autoref{sec:classes-dataflows}). Then the call graph is analyzed to identify calls to other functions (possibly residing in a remote machine), at which point functions have to be split, composing the final dataflow (\autoref{sec:splitting}). 

This dataflow graph enriched with the compiled classes, execution plans, and all metadata obtained from static analysis comprises the intermediate representation (\autoref{sec:ir}). Finally, that intermediate representation can be translated, deployed, and executed in different target systems (\autoref{sec:runtimes}). 

Note that a complete account of the analysis and transformation algorithms is not possible due to space limitations, but it will be provided in the extended version of this paper.

\begin{figure}[t!]
    \centering
    \includegraphics[width=1\columnwidth]{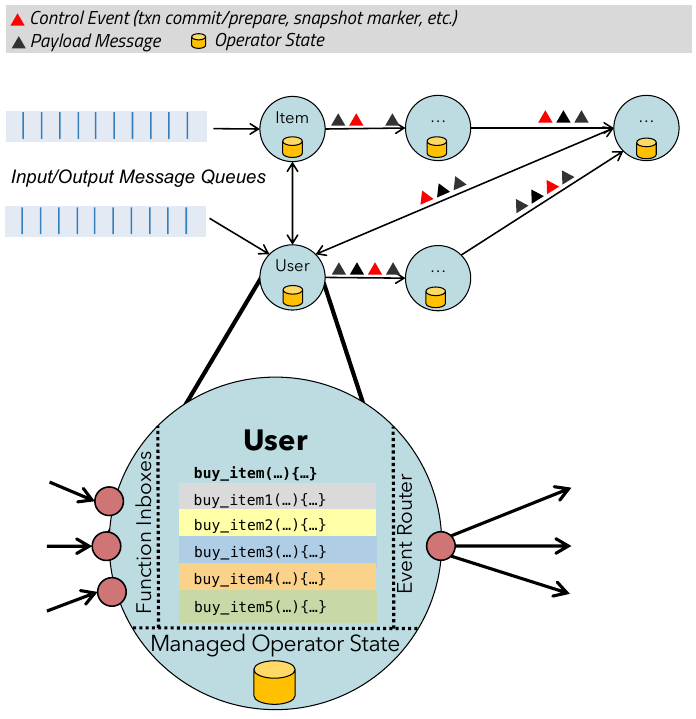}
    \vspace{-2mm}
    \caption{Logical dataflow graph of five entities, focusing on the \texttt{User} entity found in \autoref{fig:calls}.}
    \label{fig:dataflow}
    \vspace{-4mm}
\end{figure}

\subsection{Programming Model \& Limitations}
\label{sec:assumptions}

\para{Expressiveness} Our programming model allows programmers to specify simple, object-oriented Python programs. Classes can have references to other classes and call their functions. We term an instance of such a class as a \textit{stateful entity}. The \sysname{} compiler currently can analyze conditionals, \texttt{for-}loops that iterate through Python lists as well as general \texttt{while} loops. 

\para{Limitations} \sysname{} requires static type hints for the input/output of stateful entity functions; ensures the existence of those hints via a static pass over the analyzed classes. Moreover, the functions cannot be recursive. Another assumption that \sysname{} makes is that each entity contains a \texttt{key()} function. This \texttt{key()} function is used by a routing and translation mechanism to partition and distribute the load among parallel instances of that entity within a cluster. Furthermore, the key of a stateful entity cannot change throughout that entity's lifetime. Finally, the entities' state needs to be serializable, i.e., connections to databases, local pipes, and other non-serializable constructs are not allowed and will eventually generate a runtime error.

\para{Running Example}
\autoref{fig:calls} contains the code for a User and an \texttt{Item} entities. Note that since \texttt{Item} is a stateful entity, a call to \texttt{item.update\_stock(...)} is a remote function call. Both \texttt{User}, and the \texttt{Item} entities are partitioned across the cluster nodes, using the given entity's \texttt{key} function.

\subsection{From Entities to Dataflow Operators}
\label{sec:classes-dataflows}

Each Python class translates to an operator (also called a vertex) in the dataflow graph. In a dataflow graph, an operator cannot be "called" directly, like a function of an object. Instead, an \textit{event} has to enter the dataflow and reach the operator holding the \textit{code} of that entity (e.g., the \texttt{User} class) as well as the actual \textit{state} of the entities that instantiate the class (e.g., the \texttt{balance}, and \texttt{username} of the \texttt{User} in \autoref{fig:calls}).

Specifically, each dataflow operator is capable of executing all functions of a given entity and it is triggered depending on the incoming event. Since operators can be partitioned across multiple cluster nodes, each partition stores a set of stateful entities indexed by their unique key. When a function of an entity is invoked, the entity's state is retrieved from the local operator state. Then, the function is executed using the arguments found in the incoming event that triggered the call, as well as the state of the entity at the moment that the function is called.

\para{Example} A \texttt{User} operator as seen in \autoref{fig:dataflow}, is partitioned on \texttt{username}. Upon invocation of a function of the \texttt{User} entity, an event is sent to the dataflow graph's input queues. The incoming event is partitioned on \texttt{username} by an ingress router. Via the dataflow graph, the event ends up at the operator storing the state for that specific \texttt{User}. The system then reconstructs the \texttt{User} object using the operator's code and the function's state and executes the function. Finally, the function return value is encoded in an outgoing event which is forwarded to the egress router. This egress router determines if the event can be sent back to the client (caller outside the system, such as an HTTP endpoint) or needs to loop back into the dataflow in order to call another function.

\para{The need for function splitting} For simple functions that do not call other remote functions, both the translation to dataflows and the execution is straightforward. However, if the function \texttt{User.buy\_item} calls the (remote) function \texttt{item.update\_stock} whose state lies on a different partition, the situation becomes more complicated. Note that a streaming dataflow should never stop and wait for a remote function to complete and return before moving on with processing the next event. Instead, it must ``suspend'' the execution of e.g., \texttt{buy\_item} of \autoref{fig:calls}, right at the spot that the remote function \texttt{item.price()} is called until the remote function is executed and an event comes back to the \texttt{User} operator with a return value.

In order to do this, we adopt a technique to transform the imperative functions into the continuation passing style (CPS)  \cite{cps}. More specifically, we propose an approach to split a function definition into multiple ones (\autoref{sec:splitting}) at the AST level as depicted (approximately) in \autoref{fig:calls}.

\subsection{From Imperative Functions to Dataflows}
\label{sec:splitting}

\para{References to Remote Functions} After the first round of static analysis, the compiler identifies if a function definition has references to a remote stateful entity using Python type annotations. These functions may require \emph{function splitting}. The algorithm traverses the statements of a function definition and the function is split either when a remote call occurs or on a control-flow structure. For example, the following \texttt{buy\_item} calls the remote function \texttt{item.update\_stock}:
\begin{lstlisting}[style=pythonlang, numbers=left, xleftmargin=0in, label={lst:wordcount}]
def buy_item(self, amount: int, item: Item):
    total_price:int = amount * item.price
    is_removed:bool = item.update_stock(amount)
    return total_price 
\end{lstlisting}

\noindent This function is split at the assign statement on line 3 and results in two new function definitions:
\begin{lstlisting}[style=pythonlang, numbers=left, xleftmargin=0in, label={lst:wordcount}]
def buy_item_0(self, amount: int, item: Item):
    total_price:int = amount * item.price
    update_stock_arg = amount
    return total_price, {"_type": "InvokeMethod",
                         "args": [update_stock_arg], ..}
    
def buy_item_1(self, total_price, update_stock_return):
    is_removed:bool = update_stock_return
    return total_price   
\end{lstlisting}
The \texttt{buy\_item\_0} function defines the first part of the original function \textit{and} it evaluates the arguments for the remote call.
The \texttt{buy\_item\_1} function assumes the remote call \texttt{item.update\_stock} has been executed and its return variable is passed as an argument. In general, each function that was split takes as arguments the variables it references in its body and returns the variables it defines. For example, since \texttt{buy\_item\_0}  defines the variable \texttt{total\_price}, its value is returned from the function. Next, since \texttt{buy\_item\_1}  uses \texttt{total\_price}, it is defined as a parameter. 

\para{Control Flow} The compiler also needs to split functions when encountering remote function calls within control flow constructs like \textit{if}-statements or \textit{for}-loops. In short, an \textit{if}-statement is split into three new definitions: one that evaluates its conditional, one that evaluates the `true' path, and one that evaluates the `false' path. Similarly, a \textit{for}-loop is also split into three new definitions: one that evaluates the iterable, one that evaluates the \texttt{for}-body path, and one that evaluates the code path after the loop. The function splitting algorithm is recursively applied to the statements inside the \emph{for} path and inside the true and false path of the \textit{if}-statement.  

\subsection{Intermediate Representation}
\label{sec:ir}

Our intermediate representation is a stateful dataflow graph enriched with a number of aspects. After the static analysis, each dataflow operator is enriched with the entity/method names that it can run, their input/return types, as well as their method body. After splitting functions, we also need to build what we term a state machine. For every split function (Section \ref{sec:splitting}), we maintain an execution graph that tracks the execution stage of a given stateful entity's function invocation. 

Essentially, the process of deriving the state machine consists of unrolling the control flow graph of the program. Conceptually, the translation to a state machine is possible by deriving a finite representation of the program. To this end, we $i)$ do not allow unbounded recursion and we $ii)$ keep track of the current iteration for loop control structures, by enriching the state machine with the additional state. When invoking a function that was split, the state machine is inserted into the function-calling event. As the event flows through the system, the execution graph is traversed and the proper functions are called. The execution graph stores intermediate results -- the return values of the invoked functions.

\section{Runtime Dataflow Systems}
\label{sec:runtimes}

Stateful entities can be deployed as dataflow graphs to streaming dataflow systems, offering exactly-once fault-tolerance guarantees.

\para{Flink's Statefun}
The IR is translated to a streaming dataflow graph that, for example, Apache Flink can execute. In that case, a Kafka source pushes events to the ingress router, which is a map operator performing a \texttt{keyBy} operation to route an event to the correct stateful map operator instance where function execution will take place. Each execution's output is forwarded to the egress router, which forwards outputs to a Kafka sink.

We use Kafka to re-insert an event to the streaming dataflow, thereby avoiding cyclic dataflows, which are not supported by most streaming systems. Notably, our system implements all the logic required for routing and execution in this process.
On the downside, when an event reenters a dataflow to reach the next function block of a split function, race conditions attributed to events coming from non-split functions could lead to state inconsistencies due to other events changing the same function's state in the meantime. Time tracking with watermarks, support for cyclic dataflows, and locking could solve these problems. Since the IR is well-aligned with Statefun's dataflow, only simple translation and mapping is required when using the Statefun runtime.

\para{StateFlow: a transactional dataflow system} Existing dataflow systems cannot execute multi-partition transactions. To this end, we built \sysname{}, a prototype dataflow system in Python. \sysname{} treats each function -- and the state effects it creates via calls to other functions -- as a transaction with ACID guarantees. We achieve consistency by implementing an extension of Aria \cite{aria}, a deterministic transaction protocol. The dataflow system is built to allow for dataflow cycles used in function-to-function communication and leverages co-routines for optimal resource utilization. For fault-tolerance \sysname{} implements the consistent snapshots protocol~\cite{chandy1985distributed, CarboneEF17}, which has been adopted by many streaming dataflow systems~\cite{CarboneKE15, SilvaZD16, ArmbrustDT18} alongside a replayable source as an ingress, allowing \sysname{} to rollback messages and restore the snapshot upon failure. Although still a prototype, \sysname{} is already able to execute transactional workloads (YCSB-T \cite{dey2014ycsb+} and partly TPC-C) with promising performance (\autoref{sec:experiments}).

\para{Local}
A \sysname{} dataflow graph can execute all its components in a local environment. The only difference is that the state is kept in a local HashMap data structure instead of a state management backend. Local execution allows developers to debug, unit test, and validate a \sysname{} program as they would do for an arbitrary application. Afterward, they can simply deploy the program to one of the supported runtime systems.

\section{Preliminary Experiments}
\label{sec:experiments}

For the experiments of this section, we opted for running Apache Flink Statefun against \sysname{} (\autoref{sec:runtimes}). 

\para{Workload} We are using workloads A and B from the original YCSB benchmark \cite{ycsb}. A is update-heavy -- 50\% reads 50\% updates and B is ready-heavy -- 95\% reads 5\% updates. In addition, we use the transactional workload T from YCSB+T  \cite{dey2014ycsb+}, which atomically transfers an amount from one entity's bank account to another (2 reads and 2 writes). For the throughput test, we defined a mixed workload M (45\% reads 45\% updates 10\% transfers). For the latency tests, we use Zipfian and uniform key distributions. 

\para{Setup} We conducted all the experiments on 14 CPUs: 4 for the Kafka cluster, 6 for the systems, and 4 for the benchmark clients. For Statefun, we gave half of the resources to the Flink cluster and the other to the remote functions. \sysname{} requires a single core coordinator, and the rest are used for its workers.

\para{Baseline} In \sysname{}, we execute complex business logic resulting in state operations. YCSB is a benchmark that supports simple inserts, deletes, and updates, not complete executions of transactions across multiple function calls. It is therefore expected for Stateflow, since it executes function calls and application logic, to have a larger overhead than key-value stores. \sysname{} is not a key-value store; instead, it is a stateful function-as-a-service compiler and runtime that allows programmers to author object-oriented python code.

\para{Latency} In the first experiment, we measured the end-to-end latency of all the YCSB workloads against the integrated backend systems with both Zipfian and uniform key distributions at the low amount of 100RPS.
As seen in \autoref{fig:endpoint} both systems perform well with low latencies across all workloads and distributions. Some interesting observations go as follows. First, Statefun performs the same in both the A and B workloads and in both Zipfian and uniform distributions. This happens because Statefun does not use locking, allowing for concurrent access (but also inconsistency). Additionally, since all functions need to go to an external Python runtime, the cost of reads and writes are the same due to the network costs. We also observe that \sysname{} outperforms Statefun because it allows for internal function-to-function communication and does not require the roundtrips to Kafka. Note that \sysname{} additionally supports transactional workloads with higher latency than the rest but still, if we consider that a transfer operation is 2 read and 2 write operations, the transactional overhead of the system is minimal. Finally, we did not run Statefun against transactional workloads since it offers no support for transactions. 

\begingroup
\setlength{\abovecaptionskip}{0pt}
\begin{figure}[t]
    \centering
    \includegraphics[width=1\columnwidth]{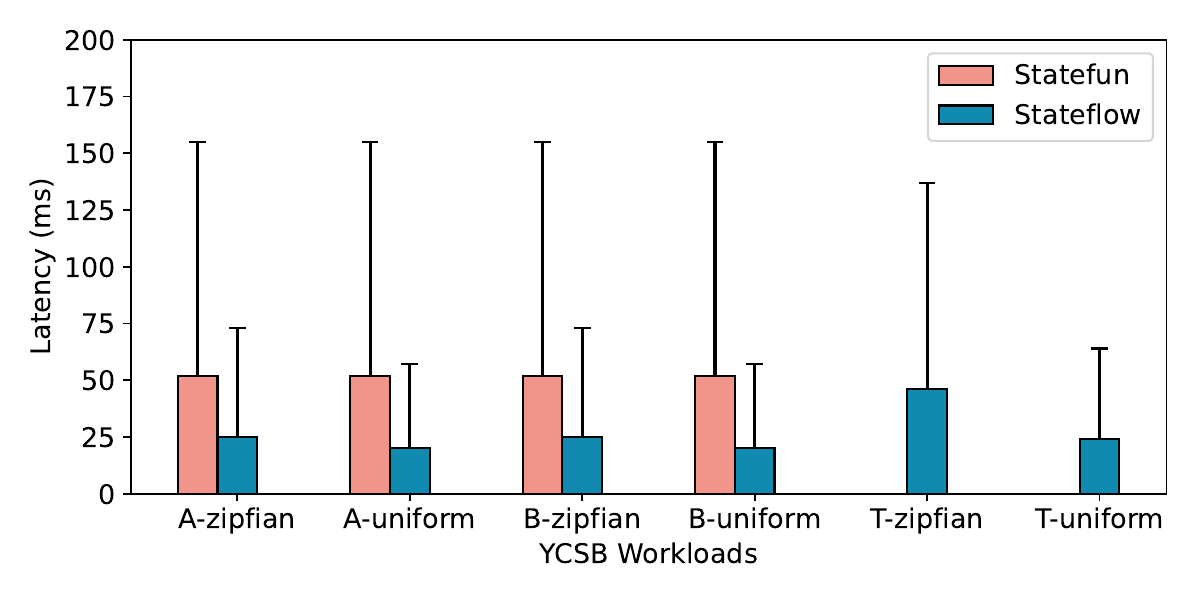}
        \caption{Average latency at the 99th percentile, in YCSB (100~RPS) with both Zipfian and uniform key distributions.}
    \label{fig:endpoint}
\end{figure}
\endgroup

\para{Throughput} In the second experiment, we gradually increase the input throughput and measure the end-to-end latency. This time we use the mixed workload that we defined, M (45\% reads 45\% updates 10\% transfers). In \autoref{fig:throughput}, we observe consistent results with the latency experiment up until the point where the difference in efficiency appears. The reason for this is that \sysname{} is using more execution cores since it bundles execution, state, and messaging. In contrast, the Statefun deployment uses half its CPUs for messaging and state within the Apache Flink cluster and the other half for execution in a remote stateless function runtime. In the current experiments this balanced deployment was the optimal one in terms of resource utilization. 

\para{System overhead} Finally, we also measured the overhead that program translation (function splits, instrumentation, etc.) incurs as part of the complete runtime (not depicted for the sake of space preservation). We created a synthetic workload in which we varied different state sizes from 50 to 200kb. For each event, we measured the duration of different runtime components. Some of the components, like object construction, are attributed to program transformation overhead, whereas others, like state storage, are attributed to the runtime. In short, function splitting/instrumentation is only responsible for less than 1\% of the total overhead.

\para{Conclusion} The experimental evaluation demonstrates the potential of dataflows as an intermediate representation and execution target for scalable cloud applications. In short, these preliminary experiments show that we can translate imperative programs that hide all the aspects of distributed systems and error management from programmers  and still achieve high performance. That said, the experiments also uncover the limitations of dataflow systems and implementation issues that we address in the following section.

\begingroup
 \setlength{\abovecaptionskip}{0pt}
 \setlength{\belowcaptionskip}{0pt}
\begin{figure}[t]
    \centering
    \includegraphics[width=1\columnwidth]{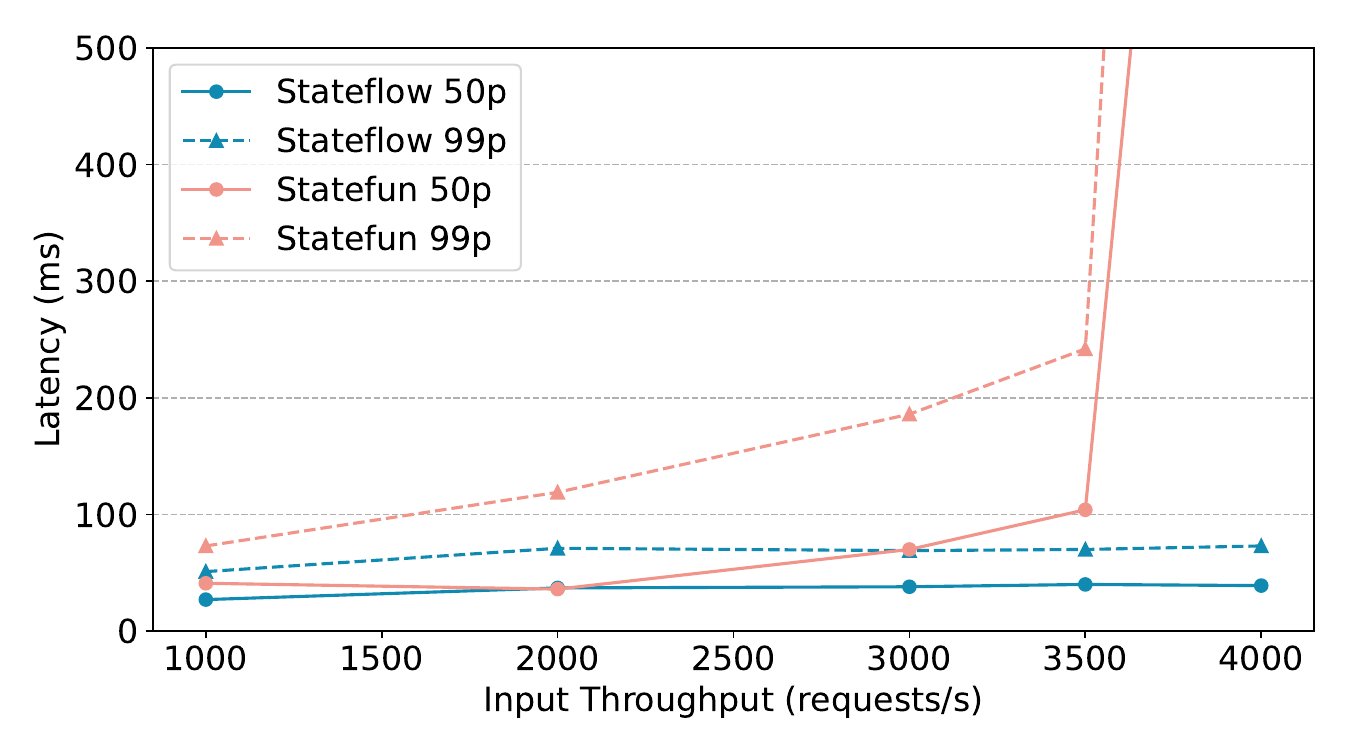}
    \caption{Average and 99th percentile latency for the M workload, with increasing input throughput.}
    \label{fig:throughput}
\end{figure}
\endgroup

\section{Open Problems \& Opportunities}
\label{sec:roadmap}

The ability to query the global state of a dataflow processor, as well as perform transactional state updates on its state, can transform a dataflow processor into a full-fledged, distributed database system. The envisioned system will be capable of executing Turing-complete ``stored procedures''  (such as the entity functions in the case of this paper) that are distributed, partitioned and can perform function-to-function calls with exactly-once guarantees. This is the ultimate goal of this work.

In this section, we discuss a number of opportunities emerging mainly from transactional workloads with low-latency requirements and outline future research directions to enable the adoption of dataflow systems for executing general cloud applications.

\para{Program Analysis}
The dataflow model is essentially a {\it finite state machine} where nodes are the functions from the original ({\it Turing-complete}) program and arcs indicate event flow. 
In the case of loops, events also carry information about the previous iterations of the loop (e.g., the variables that are read and written in the loop body and in the loop condition clause). This information handles loops correctly (\autoref{sec:ir}).
For method calls, if a method is mapped to a single state, it would be problematic to determine where to return after a call if in the codebase there are multiple calls that have different return points. We map each method {\it call} into a transition to a state that is specific for that call. This means that calls to the same method may result in a different state in the automata, ensuring that each of these states has as a next state the correct return point. This approach requires to {\it unroll} the program, expanding each potential method call that may occur at runtime into a different state.

Following this approach, recursive functions would result in a state for each recursive step. Since unbounded recursion would result in infinite automata, we prohibit recursion.
Yet, from a compiler perspective, since a program can be CPS-transformed, recursion can be translated into loops via tail-call elimination~\cite{10.1145/214448.214454}, which could potentially affect the dataflow engine's performance.

In addition, in what is traditionally referred to as {\it dataflow languages} (e.g., Esterel~\cite{BERRY199287}, Lucid~\cite{lucid85}), the computation is driven by data propagation -- just like in streaming dataflows. However, the expressivity of such languages has been intentionally limited to enable efficient execution (automatic) verification techniques. While in this work we aim to target Turing-complete Python programs, the trade-off between expressivity, efficiency, and automatic verification is yet to be researched in the future.

\para{Transactions} Current dataflow systems guarantee the consistency of single-event effects on a given key of the state.  In order to support transactional executions across stateful entities, we could employ \textit{single-shot} transactions \cite{one-shot-h-store} or, like in our prototypical dataflow system (\autoref{sec:runtimes}), borrow ideas from deterministic databases \cite{Thomson2010Calvin,Abadi2018Deterministic,aria} for minimizing the coordination of transactions. In practice, a large percentage of transactions can be expressed as single-shot transactions \cite{single-shot-aws}; very popular databases such as Amazon's DynamoDB \cite{sivasubramanian2012amazon} and VoltDB \cite{stonebraker2013voltdb} support single-shot transactions. These ideas can define how a  programming model can support patterns that have been adopted by practitioners in the last years, starting from SAGAs \cite{sagas} and Try-Confirm-Cancel \cite{helland2016life}.

\para{Exactly-once, Latency \& External Systems} Exactly-once guarantees can incur high latency: the outputs of a dataflow only become visible after an epoch terminates successfully \cite{CarboneEF17}. Epoch intervals cannot be too small because they would incur a high overhead. However, one can leverage causal recovery \cite{wang2019lineage} and determinants \cite{silvestre2021clonos} alongside  replayable sinks in order to minimize the latency within each epoch.
The replayable sinks are required  to be able to retrieve determinants. However, at the border of a system, i.e., when a message leaves the dataflow graph and is sent to an external system, replayable sinks may be hard to assume. In that case, one should make use of more traditional techniques for deduplication (e.g., the common idempotence keys used in the HTTP protocol). Under certain assumptions (deterministic computations, persistent/replayable request queues, etc.), such idempotence keys can be generated automatically. However, this will not be the case for a generic distributed application, which will have to generate, keep track of, check, and recycle unique identifiers to enforce the delivery of its output exactly-once. These issues have not been studied enough in the context of distributed databases, neither in models for cloud programming.

\para{Querying Stateful Entities}
In previous work \cite{squery}, we have shown that querying the global state of a dataflow processor can be, not only efficient but can also come with certain correctness guarantees. Some work on querying actors has already been done in the context of Orleans \cite{bernstein2017indexing}. However, querying (e.g., with SQL) a set of entities still poses a number of challenges, especially with respect to the tradeoff between the freshness and consistency of query results. To this end, we could borrow ideas from RAMP (read-atomic) transactions \cite{bailis2016scalable} that match well the execution model of transactions and read operations in stateful entities.

\section{Related Work}

The idea of democratizing distributed systems programming is not new. For instance, in~\cite{newdirectionscloud},  the authors mention that a combination of dataflows and reactivity would provide a good execution model for cloud applications. In this work, we share the same belief and build a prototype towards that direction.

\para{Programming models} In the past, approaches like Distributed ML~\cite{krumvieda1993distributed}, Smalltalk~\cite{deutsch1984efficient}, and Erlang~\cite{armstrong2013programming} aimed at simplifying the programming and deployment of distributed applications. Many of those ideas, including the Actor model, can be reused and extended today.
Erlang implemented a flavor of the actor model. 
Akka~\cite{wyatt2013akka} offers a low-level programming model for actors. Closest to our work is the Virtual Actors model introduced by Orleans \cite{bykov2011orleans,bernstein2014orleans}, which aims at simplifying Cloud programming and even supports some form of transactions~\cite{eldeeb2016transactions}. 
However, Orleans requires a specialized runtime system for virtual actors, which does not support exactly-once messaging and does not compile its actors into stateful dataflows. Nonetheless, our work is heavily inspired both by Orleans and by Pat Helland's entities \cite{helland2016life}.

\para{Imperative programming to Dataflows} The idea of translating imperative code to dataflow is not new. In the database community, there has been work on detecting imperative parts of general applications that can be converted into SQL queries (e.g., \cite{emani2017dbridge}) but also for automatic parallelization of imperative code in multi-core systems. For instance, the work by Gupta and Sohi~\cite{gupta2011dataflow} compiles sequential imperative code to dataflow programs and executes them in parallel. Our work draws inspiration from both these lines of work and extends them by taking into account the partitioning of state as well as other considerations that we outline in~\autoref{sec:roadmap}.

\para{Stateful Functions} A new breed of systems marketed as stateful functions such as Cloudburst \cite{cloudburst}, Lightbend's \url{Cloudstate.io} and Apache Flink's \url{Statefun.io}~\cite{statefun}, as well as our early prototype in Scala \cite{AkhterFK19} also aim at abstracting away the details of deployment and scalability. However, none of those compiles general-purpose object-oriented code into dataflows.

\section{Conclusions}

In this vision paper we argue that if we want to hide failures from the top-level programming models of Cloud applications, exactly-once guarantees should become a first-class citizen. While dataflow systems can provide such guarantees, their programming model makes the development of general Cloud applications cumbersome. To this end, we have developed a compiler pipeline that statically analyzes an object-oriented Python application in order to create an intermediate representation in the form of a dataflow graph, and then submit that dataflow graph to existing dataflow systems. Leveraging dataflow systems' exactly-once guarantees can essentially hide all Cloud failures from programmers with low overhead: our preliminary experimental evaluation demonstrates that function splitting and program transformation incur less than 1\% overhead and the YCSB+T benchmark, with low-latency execution. 

\para{Current Status} Despite the encouraging results, lots of problems remain open: specifically in the area of transaction execution, programming models, program analysis, and dataflow engines for general cloud applications. Our work currently focuses primarily on $i)$ strengthening the formal underpinnings of program transformation to dataflows, $ii)$ extending the programming model with different transactional paradigms, and $iii)$ further developing \sysname{}, our novel transactional dataflow system.

\vspace{5mm}

\begin{acks}
This publication is part of project number 19708, of the Vidi research program which is partly financed by the Dutch Research Council (NWO).

\vspace{5mm}

\noindent In memory of Eelco Visser.

\vspace{5mm}
\end{acks}

\bibliographystyle{ACM-Reference-Format}
\bibliography{main}

%
% ...

\end{document}